\documentstyle[epsfig,12pt]{elsart}
\begin{document}
\newcommand{\kp}{K^+}
\newcommand{\gk}{\vec{\gamma}\vec{k}}
\newcommand{\gE}{\gamma_0 E_k}
\newcommand{\ppl}{\vec{p}}
\newcommand{\bcm}{\vec{b}^{\star}}
\newcommand{\becm}{\vec{\beta}^{\star}}
\newcommand{\bepl}{\vec{\beta}}
\newcommand{\rcm}{\vec{r}^{\star}}
\newcommand{\rpl}{\vec{r}}
\newcommand{\A}{{$\mathcal A$}}
\newcommand{\wpk}{ \omega_{p-k}}
\newcommand{\Journal}[4]{ #1 {\bf #2} (#4) #3}
\newcommand{\NPA}{Nucl.\ Phys.\ A}
\newcommand{\PLB}{Phys.\ Lett.\ B}
\newcommand{\PRC}{Phys.\ Rev.\ C}
\newcommand{\ZPC}{Z.\ Phys.\ C}
\newcommand{\be}{\begin{equation}}
\newcommand{\ee}{\end{equation}}
\begin{frontmatter}

\title{Isospin Effects on Strangeness in Heavy-Ion Collisions}

\author[auth]{V. Prassa}
\author[giessen]{T. Gaitanos}
\author[catania]{G.Ferini}
\author[catania]{M. Di Toro}
\author[auth]{G.A. Lalazissis}
\author[muenchen]{H.H. Wolter}

\address[auth]{Department of Theoretical Physics,
Aristotle University of Thessaloniki, Thessaloniki Gr-54124,Greece}
\address[giessen]{Institut f\"{u}r Theoretische Physik,
Justus-Liebig-Universit\"{a}t Giessen,
D-35392 Giessen, Germany}
\address[catania]{Laboratori Nazionali del Sud INFN, I-95123 Catania,
Italy}
\address[muenchen]{Sektion Physik, Universit\"at M\"unchen,
D-85748 Garching, Germany}
\address{email: Theodoros.Gaitanos@theo.physik.uni-giessen.de}
\begin{abstract}
Kaon properties are studied within the framework of a fully covariant
transport approach. The kaon-nucleon potential is evaluated in two schemes,
a chiral perturbative approach and an effective One-Boson-Exchange model.
Isospin effects are explicitly accounted for in both models. The transport
calculations indicate a significant sensitivity of momentum distributions
and total yields of $K^{0,+}$ isospin states on the choice of the
kaon-nucleon interaction. Furthermore, isospin effects are rather moderate on
absolute kaon yields, but appear on strangeness ratios.
This is an important issue in determining the high density symmetry energy
from studies of strangeness production in heavy-ion collisions.
\end{abstract}
\end{frontmatter}

\date{\today}

\section{Introduction}

The knowledge of the in-medium hadronic properties at supra-normal
densities is of major importance for the understanding of (nuclear)
astrophysics such as the physical mechanism of supernovae explosions
and the physics of neutron stars \cite{astro}.
Strangeness production at intermediate energy heavy-ion collisions has been
a very helpful tool in studying the hadronic equation of state (EoS) and
the properties of strangeness matter under extreme conditions of baryon
density and temperature \cite{Fang,fuchs06}.

It is well established that ($K^{0,+}$) kaons feel a weak repulsive potential
of the order of $20-30$ MeV at saturation density $\rho_{sat}=0.16~fm^{-3}$,
as it has been verified by studying strangeness flows in heavy-ion collisions
\cite{fuchs06,larionov}. The determination of the high density behavior
of the nuclear EoS has been successfully investigated in this
context \cite{ako,kaos,fuchs}.
It has been realized that strangeness production could also serve as 
 an important messenger
for the isovector part of the nuclear EoS at high densities, i.e. for
the symmetry energy \cite{ferini2}. A reliable determination of the symmetry
energy at supra-normal densities from heavy-ion collisions is still an object
of current debates and investigations \cite{others,ferini,BALI_pi}.

However, a reliable study of the high density nuclear EoS and, in particular,
its isovector part involves a detailed analysis of the role of the isospin
dependent part of the kaon-nucleon interaction on strangeness production in
intermediate energy nucleus-nucleus collisions, which has not been
studied so far. Theoretical studies on static hadronic matter
based on the Chiral Perturbation Theory (ChPT) \cite{kaonpot} and on the
One-Boson-Exchange approach (OBE) \cite{idkpot} predict a splitting of
self energies between positive charged ($K^{+}$) and neutral ($K^{0}$)
kaons. This isospin effect increases with baryon density and isospin asymmetry,
and it may thus influence the production and propagation of particles with
strangeness in dense hadronic matter during a nucleus-nucleus collision.
It may thus be a sensitive observable to determine the symmetry energy at high densities.

This work is an extension of former theoretical investigations of kaon 
medium effects in heavy-ion collisions \cite{Fang} to asymmetric hadronic 
matter. The main topic of our study is therefore the influence
of the isovector sector of the kaon-nucleon interaction on kaon dynamics
at incident energies near the strangeness production 
threshold. In extension to our previous work \cite{prassa1}, two different models for the
kaon-nucleon interaction and their particular isovector properties
are now discussed in detail: one based on the Chiral Perturbation Theory
(ChPT) \cite{kaonpot} and a second one based on the One-Boson-Exchange (OBE)
approach for the effective meson-nucleon field theory \cite{idkpot}.
We use a fully relativistic transport model, in which isospin effects are
included in the mean-field and in the collision integral \cite{ferini},
and also in the different realizations of the kaon potential.
These models are applied in heavy-ion collisions at intermediate energy below
the threshold for strangeness production. The production of the
$(K^{+},K^{0})$ isospin states and, in particular, their ratio is extensively investigated
within the different scenarios for the isovector sector of the nuclear mean-field
and the kaon-nucleon potential. The antikaon field
$\overline{K}=(K^{-},\overline{K^{0}})$ is not considered here. A discussion
of the stronger and more complex $\overline{K}$-nucleon interaction would go
beyond the scope of this work, e.g., we refer to Refs. \cite{kolos}.

\section{Theoretical description}

In this section we describe the theoretical model used in this work. After a brief
introduction of the transport equation we focus on the different treatments of
the kaon-nucleon potential in isospin-asymmetric hadronic matter, which is the major
focus of the present work.

\subsection{The transport equation}

The theoretical description of a heavy ion collision is based on the
relativistic Boltzmann-Uehling-Uhlenbeck (RBUU) dynamical approach,
as introduced in Refs. \cite{fuchs_RLV,giessen,ko} and then extended for
isospin and strangeness degrees of freedom \cite{ferini}:
\begin{eqnarray}
& & \left[
k^{*\mu} \partial_{\mu}^{x} + \left( k^{*}_{\nu} F^{\mu\nu}
+ m^{*} \partial_{x}^{\mu} m^{*}  \right)
\partial_{\mu}^{k^{*}}
\right] f_i(x,k^{*}) = {\cal I}_{\rm coll}
\label{rbuu}
\quad .
\end{eqnarray}
The RBUU transport equation describes the phase space evolution of the
one-particle density distribution functions $f_i(x,k^{*})$ for the
different particle species under consideration
under the influence of a mean-field (lhs of eq.(\ref{rbuu})) and binary collisions
(rhs of eq.(\ref{rbuu})) which couple the different fields. 
Here we consider the propagation of all baryons up 
to the $\Delta$-resonance and hyperons ($\Lambda,\Sigma$). The production of
pions results through the resonance decay and kaons ($K^{0,+}$) are produced
in baryon-baryon and baryon-pion binary collisions (see for details
Ref. \cite{prassa1}). The mean-field or the nuclear EoS enters via the effective mass
$m^{*}=M-\Sigma_{s}$, the kinetic momenta $k^{*\mu}=k^{\mu}-\Sigma^{\mu}$
and the field tensor
$F^{\mu\nu} = \partial^\mu \Sigma^\nu -\partial^\nu \Sigma^\mu$,
where $\Sigma_{s}$ and $~\Sigma^{\mu}$ are the Lorentz-scalar and
Lorentz-vector components of the baryon self energy. Pions are propagated under
the influence of the Coulomb field only. Mesons with strangeness obey the same transport
equation (\ref{rbuu}), however, with different self energies with respect to the baryon
ones, as it will be discussed below. The collision integral is determined
by the total elastic and inelastic in-medium cross sections and includes
all important channels on pion and kaon production, without the inclusion of
antikaons. Isospin effects on the threshold conditions are explicitly accounted for
to respect energy conservation \cite{ferini}. More details can be found in
Ref. \cite{ferini,prassa1}. Furthermore, the quasi-elastic $KN\leftrightarrow KN$
channel with isospin exchange is included, however, it does not essentially influence the
strangeness ratios.
\subsection{The $ChPT$ potential in hadronic matter}
Kaplan and Nelson were the first to apply the chiral Lagrangian approach to
the properties of kaons in dense nuclear matter \cite{kaonpot}. Starting
from the effective chiral Lagrangian an
in-medium Klein-Gordon equations for kaons at the mean field
level follow can be derived
\begin{equation}
\left[\partial_\mu\partial^\mu + \frac{3i}{4f_{\pi}^{*2}}j_{\mu}\partial^{\mu} +
\left(m_{K}^{2}- \frac{\Sigma_{KN}}{f_{\pi}^{2}}\rho_{s}\right)\right] \phi_{K}(x)= 0
\label{kaonmean}
\quad ,
\end{equation}
where $j_{\mu}$ and $\rho_{s}$ are the baryon current and the scalar density,
respectively.
Eq. (\ref{kaonmean}) can be rewritten in a more compact form as
\begin{equation}
\left[\left(\partial_{\mu} + iV_{\mu} \right)^{2} +  m_{K}^{*2}\right]\phi_{K}(x)= 0 \quad .
\label{kaonmean2}
\end{equation}

In isospin asymmetric matter the kaon effective mass is given by
\cite{idkpot}
\begin{equation}
m_{K}^{*2} =m_{K}^{2} -\frac{\Sigma_{KN}}{f_{\pi}^{2}}\rho_{s} -
\tau_{3}\frac{C}{f_{\pi}^{2}}\rho_{s3}+ V_{\mu} V^{\mu} 
\label{kaoneffiso}
\quad ,
\end{equation}
where $\rho_{s},~\rho_{s3}$ are the total and isospin scalar baryon
densities. The vector potential is defined as
\begin{equation}
V_{\mu} = \frac{3}{8f^{*2}_{\pi}}j_{\mu} +\tau_{3}
\frac{1}{8f^{*2}_{\pi}}j_{\mu3} 
\label{kapotiso}
\quad ,
\end{equation}
with $j^{\mu},~j^{\mu3}$ isoscalar and isovector baryon currents,
respectively, and $\tau_{3}=\pm~1$ for $K^{+}$ and $K^{0}$.
$m_{K}=494$ $MeV$ is the free kaon mass, $f_{\pi}=93$ $MeV$ the pion decay
constant and  $\Sigma_{KN}$ the kaon-nucleon sigma term, usually chosen in
the range $350\div 450$ $MeV$.
$f^{*}_{\pi} \simeq 0.6 f^{2}_{\pi}$ is the in-medium pion
decay constant \cite{BroRho} which appears only in the vector field.
According to Brown and Rho \cite{BroRho} this accounts for the fact that the
enhancement of the the scalar part is compensated by higher
order corrections in the chiral expansion. Finally, the parameter $C$ is fixed to
the value of $33.5 \, MeV$ \cite{idkpot}.

\subsection{Effective mean field potential in hadronic matter}
In an alternative approach
kaon potentials can also be derived within an effective meson field
$OBE$ approach  \cite{idkpot}. In this picture the
kaons interact through the exchange of mesons with different Lorentz
properties ($\sigma,~\omega,~\rho,~\delta$). These mesons mediate the
isoscalar-scalar, isoscalar-vector, isovector-vector and
isovector-scalar parts of the kaon-kaon interaction.
In the mean-field approximation the kaon equation of motion reads
\begin{eqnarray}
& &\left[
\partial^{\mu}\partial_{\mu}+m_{K}^{2}+g_{\sigma K}m_{K}\sigma+
\frac{g_{\delta K}g_{\delta N}}{m_{\delta}^{2}}m_{K}\tau_{3}\rho_{s3}
 + 2\left(\frac{g_{\omega K}g_{\omega N}}{m_{\omega}^{2}}j_{\mu}+
\frac{g_{\rho K}g_{\rho N}}{m_{\rho}^{2}}\tau_{3}j_{\mu 3}\right)i\partial^{\mu}
\right.
\nonumber\\
& &
\left.
- \left(\frac{g_{\omega K}g_{\omega N}}{m_{\omega}^{2}}j_{\mu}
 +\frac{g_{\rho K}g_{\rho N}}{m_{\rho}^{2}}\tau_{3}j_{\mu 3}\right)^{2}
\right]\phi_K(x)=0
\label{kaoneq2}
\quad .
\end{eqnarray}
As in the ChPT case (see Eq. (\ref{kaonmean2})) Eq. (\ref{kaoneq2}) can be
written more compact as:
\begin{equation}
\left[(\partial^{\mu}+ iV^{\mu})^{2}+m_{K}^{*2}\right] \phi_K(x)=0
\label{kaoneq}
\end{equation}
The vector potential $V^{\mu}$ and the effective mass $m_{K}^{*}$  are now
defined differently relative to the ChPT approach
\begin{eqnarray}
V^{\mu}& = & \frac{1}{3}\left(f_{\omega}^{*}j^{\mu}+\tau_{3} f_{\rho}j^{\mu}_{3}\right)
\label{vpot2}\\
m_{K}^{*2}& = & m_{K}^{2}-\frac{m_{K}}{3}\left(g_{\sigma N}\sigma+\tau_{3}
f_{\delta}\rho_{s3}\right)
\label{emass}
\quad ,
\end{eqnarray}
where $j^{\mu}=j^{\mu}_{p}+j^{\mu}_{n}$ and $j^{\mu}_{3}=j^{\mu}_{p}-j^{\mu}_{n}$,
$\sigma$ and $\rho_{s3}$
are the isoscalar, isovector baryon currents, the isoscalar-scalar field
and isovector-scalar baryon density, respectively. 
We will use a simple constituent quark-counting 
prescription to relate the kaon-meson couplings to the nucleon-meson
couplings, i.e. a reduction by a factor $3$; $g_{iK}=\frac{1}{3}g_{iN}$
($i=\sigma,\omega,\rho,\delta$). Thus, in the isoscalar-vector sector we
have $f_{\omega}^*=1.4(\frac{g_{\omega N}}{m_{\omega}})^{2}$
due to the enhanced kaon-scalar/vector coupling. 
In the isovector-vector sector one obtains
$f_{\rho}=(\frac{g_{\rho N}}{m_{\rho}})^{2}$. The scalar kaon field
finally includes the isoscalar-scalar and isovector-scalar couplings
$g_{\sigma N}$ and $f_{\delta}=(\frac{g_{\delta N}}{m_{\delta}})^{2}$,
respectively. For the parameters of the nucleon-meson couplings
we refer to Ref. \cite{ferini}.
We note that the picture of a simple constituent quark-counting may not
be consistent with $KN$ scattering in free space. However, this
is not of importance here, since $KN$ scattering may influence kaon spectra,
but not absolute kaon yields and their ratio, which is the main topic of
this work.

\begin{figure}[t]
\unitlength1cm
\begin{picture}(8.,5.)
\put(0.5,0.0){\makebox{\epsfig{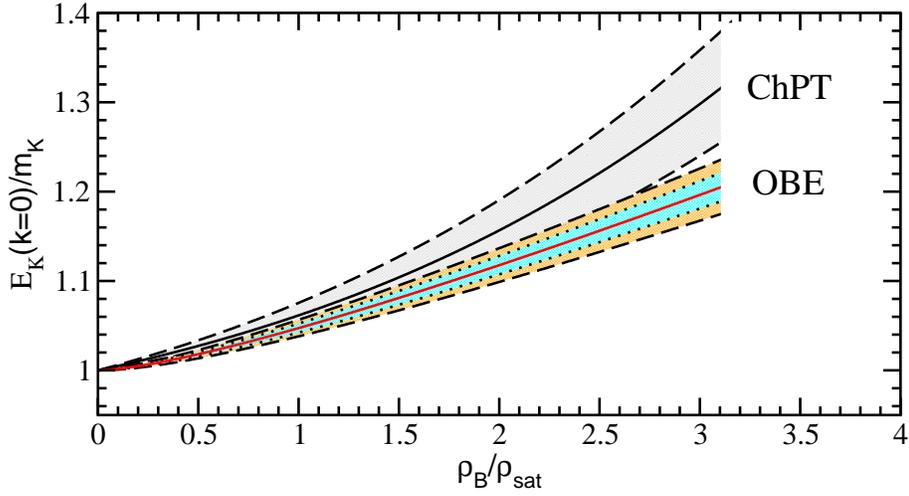}}}
\end{picture}
\caption{Density dependence ($\rho_{sat}$ is the saturation density)
of the in-medium kaon energy in units of the free kaon mass
($m_K=0.494~GeV$). Upper solid curve refers to symmetric matter in the
$ChPT$ model, while the band gives the isospin splitting between
$K^{0}$ and $K^{+}$ (upper and lower dashed curves of the band, respectively)
for an asymmetry of $\alpha=0.2$. The solid lower curve refers
to the $OBE$ approach for symmetric matter. The isospin
splitting between $K^{0}$ (upper dashed and dotted lines) and $K^{+}$
(lower dashed and dotted lines) originating from the model with only a
$\rho$-meson ($NL\rho$) is indicated by the inner band, while the outside
band refers to the model with both, the $\rho$ and $\delta$ mesons ($NL\rho\delta$).
The asymmetry parameter is $\alpha=0.2$.
}
\label{kmass}
\end{figure}

Comparing both models for the kaon-nucleon interaction,
Eqs. (\ref{vpot2},\ref{emass}) versus (\ref{kaoneffiso},\ref{kapotiso}),
one sees that in the $OBE$ model the nucleon-meson couplings enter
explicitly in the kaon self energies, while in the $ChPT$ approach
the kaon mean field is determined by the parameters $\Sigma_{KN}$ and
$C$. It is therefore helpful to study the isospin dependence of the
$OBE$ model for various parametrizations of the isovector EoS, i.e.,
the symmetry energy. As in previous works \cite{ferini,prassa1}, we
use the $NL$, $NL\rho$ and $NL\rho\delta$ prescriptions for the isovector
mean-field. In the first one ($NL$) no isospin potential is included.
$NL\rho$ contains only the $\rho$ meson, while $NL\rho\delta$ accounts
for both, the $\rho$ and the
$\delta$ mesons in the isovector potential. The symmetry energy becomes
stiffer in the direction $NL\to NL\rho \to NL\rho\delta$, since one has
to increase the $\rho$-meson coupling when the $\delta$ meson field is
taken into account in the potential part of the symmetry energy.

The crucial quantity for the interpretation of the transport results
shown later is the kaon in-medium energy, given by the dispersion relation:
\begin{equation}
E_K({\bf k}) = k_0 = \sqrt{{\bf k}^{2} + m^{*2}_{K}}+ V_{0}
\label{energy}
\end{equation}
The density dependence of the in-medium kaon energy
$E_{K}({\bf k}=0) = m^{*}_{K} + V_{0}$ for $K^{0,+}$ is shown in Fig.
\ref{kmass} for both models for the kaon-nucleon potential. In general the
$ChPT$ approach leads to a stiffer density dependence of the in-medium kaon
energy relative to the $OBE$ model. This feature is more pronounced at
high densities $\rho\sim (2-3)\rho_{sat}$, which are reached in intermediate
energy heavy-ion collisions near the kaon production threshold. Thus, one expects
observable effects of the model of the symmetry energy 
in the produced kaons ($K^{0,+}$) from heavy-ion collisions.

The isospin effect for the considered asymmetry parameter of
$\alpha=\frac{N-Z}{N+Z}=0.2$ ($N,Z$ are the neutron and proton numbers), which
corresponds approximately to ${}^{197}Au$-nuclei, is pronounced at the baryon densities
of $\rho\sim (2-3)\rho_{sat}$, see again Fig.~\ref{kmass}. According to
Eqs. (\ref{kaoneffiso},\ref{emass}) the isovector contributions to the
kaon-nucleon $ChPT$ and $OBE$ scalar self energies lead to an effective mass
splitting between positive ($K^{+}$) and neutral ($K^{0}$) kaons. Also the vector
self energies between $K^{+}$ and $K^{0}$ differ according
Eqs. (\ref{kapotiso},\ref{vpot2}). These isospin effects finally lead to the pronounced
isospin dependence of the in-medium kaon energy. We should note here the different
origin of the isospin part of the kaon-nucleon potential between the two approaches.
In the $ChPT$ case the isospin dependence enters only through the isospin-currents,
while in the $OBE$ model the effective isovector meson-nucleon couplings are
explicitly included. This leads to an isospin splitting between $K^{0}$ and $K^{+}$,
which additionally depends on the  model for the symmetry energy.

The non-trivial isospin dependence of the two models for the kaon-nucleon potential
may have interesting effects on the absolute $K^{0}$ and $K^{+}$ yields and their
collective flows and, in particular, on the strangeness ratio $K^{0}/K^{+}$ in
heavy-ion collisions at intermediate energies near the kaon production threshold,
which is the topic of the next section.

\section{Results}

The results of the following dynamical calculations are based on the
transport equation (\ref{rbuu}), in which the isospin effects of the
kaon-nucleon potentials are of particular interest, since they are
investigated for the first time and may affect absolute kaon yields and
strangeness ratios.

We first discuss in subsection $3.1$ the qualitative features of the two models of the
kaon-nucleon potential and compare with experiments.
For a transparent discussion on the the isospin effects
the following strategy has been chosen: the isoscalar part of the nuclear
mean-field is the same for all different cases of transport calculations,
while the isovector channel includes only the $\rho$ meson. In the strangeness
sector following cases of different transport calculations will be discussed:
\begin{enumerate}
\item $ChPT$ refers to transport results with only the isospin independent part of the 
 chiral model for the kaon potential.
\item $OBE$ refers to transport results with only the isospin independent part of the 
 effective mean-field model for the kaon potential.
\item $IChPT$ refers to transport results with the full isospin dependent chiral model for the
kaon potential.
\item $IOBE$ refers to transport results with the full isospin dependent effective mean-field
model for the kaon potential.
\end{enumerate}
Thus in the first two cases isospin effects are included only in the baryonic sector
($\rho$ meson), while in the last two ones they are taken into account 
also in the strangeness sector.

We then investigate in subsection $3.2$ in detail the role of the different isospin-dependent
kaon-nucleon potentials on the strangeness $(K^{0}/K^{+})$-ratio in $Au+Au@1~AGeV$
central collisions. Here we particularly distinguish for each case listed above
between the $NL$, $NL\rho$ and $NL\rho\delta$ models, in which the symmetry energy
becomes more stiffer in the sequence $NL\to NL\rho\to NL\rho\delta$.

\subsection{Kaon yields}

\begin{figure}[t]
\unitlength1cm
\begin{picture}(8.,7.)
\put(2,0.0){\makebox{\epsfig{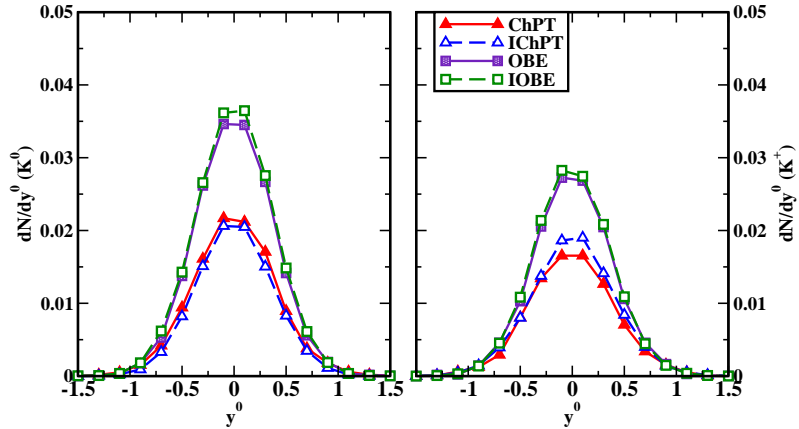}}}
\end{picture}
\caption{ Rapidity distributions of the $K^{0}$ (left panel)
and $K^{+}$ (right panel) yields for central ($b=0fm$) $Au+Au$-collisions
at $1~AGeV$. Calculations with $ChPT$ (solid, full triangles),
with $IChPT$ (dashed, open triangles), with $OBE$ (solid, full squares)
and with $IOBE$ (dashed, open squares) kaon potential are shown. In the
calculations with the isospin dependence only the $\rho$-meson ($NL\rho$)
has been used.
}
\label{kpotrap}
\end{figure}

We start the discussion with the influence of the different kaon potentials
on absolute $K^{+,0}$ meson yields. This is shown in
Fig. \ref{kpotrap} for heavy-ion collisions between heavy systems (Au+Au),
 in terms of their rapidity distributions for the four cases.
Consistent with Fig.~\ref{kmass} the transport calculations with the $ChPT$ model
with the larger repulsion
leads to overall smaller kaon yields with respect to
those in the $OBE$ approach. The inclusion of the isospin dependent
part in the kaon potentials (compare the curves $IChPT$ and $ChPT$ as
well as those between $IOBE$ and $OBE$ in Fig. \ref{kpotrap})
changes only very little the absolute yields. This is again consistent with
Fig. \ref{kmass}, where the isospin effects on the two isovector states are small
at baryon densities around $\rho_{B}\sim 2\rho_{sat}$. Furthermore, in
Fig. \ref{kmass} the asymmetry parameter was fixed to a value of $\alpha=0.2$
corresponding to the asymmetry of a ${}^{197}Au$-nucleus), while
in heavy-ion collisions the local isospin-asymmetry, in the stage
of kaon production, is reduced with respect to that of the initial situation
\cite{ferini2}. This is because of  partial isospin equilibration due to stopping
and inelastic processes with associated isospin exchange.

\begin{figure}[t]
\unitlength1cm
\begin{picture}(8.,7.)
\put(2.,0.0){\makebox{\psfig{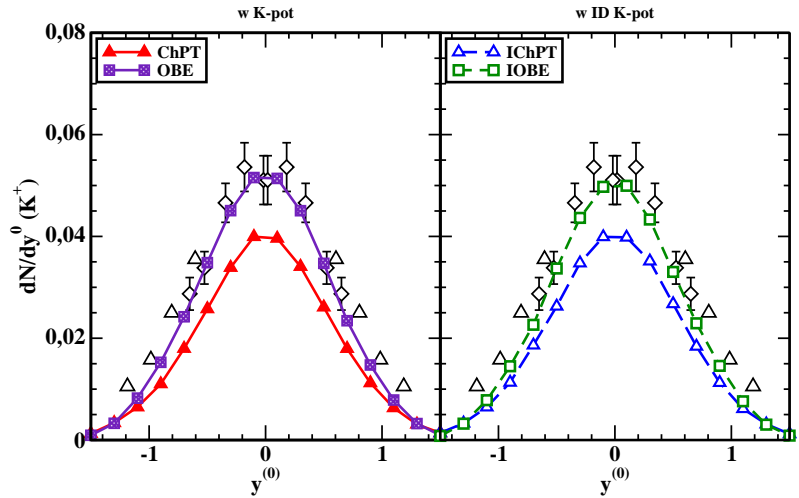}}}
\end{picture}
\caption{ $K^{+}$ rapidity distributions for
semi-central ($b<4~fm$) Ni+Ni reactions
at 1.93 AGeV. Theoretical calculations (as indicated) are compared with
the experimental data of FOPI (open triangles) and KaoS (open diamonds)
collaborations \protect\cite{expf,expk}.
}
\label{kNiNi}
\end{figure}

It is interesting to study the role of the different kaon potentials in
lighter colliding systems, such as $Ni+Ni$, in which also experimental
information is available \cite{expf,expk}. Fig.~\ref{kNiNi} shows the
comparison between the data and the transport calculations obtained with
both sets of the kaon potential without and with the isospin dependence
(left and right panel, respectively). The isospin dependence is again 
not of relevance here. The different kaon potentials influence the kaon
yields in a similar manner as in the previous figure. These in-medium
dependencies are less pronounced now, since less compression is achieved in
the stage of kaon production for the lighter $Ni+Ni$-system.

Both sets of calculations are close to the experimental
data, with the tendency of an improvement in the $OBE$ and $IOBE$ cases.
However,
this result does not neccesarily mean that the $OBE$ and $IOBE$ kaon-nucleon
potentials are the most realistic ones, because other physical quantities
may influence the transport calculations. A general and non-trivial
feature in heavy-ion collisions concerns the treatment of the in-medium
effects on the elastic and inelastic cross sections, which are usually
parametrized to empirical free NN-scattering. Microscopic studies
on elastic NN-scattering at finite baryon density exist \cite{DBx},
but for inelastic processes, e.g., $NN\to N\Delta$, theoretical calculations
are very rare and often limited in baryon density and momentum \cite{DBx}.
Attempts have been recently done in studying this non-trivial problem in
heavy-ion collisions \cite{prassa1} using a simple parametrization from
microscopic Dirac-Brueckner calculations \cite{DBx}.
For the strangeness sector, e.g., $BB\to BYK$ and $\pi B\to YK$ ($B$, $Y$
and $K$ stand for nucleons or higher resonances, hyperons and kaons,
respectively), the in-medium effects are trivially included only in the
phase-space factors and the threshold conditions,
but not in the matrix elements,
i.e., here free parametrizations are
used \cite{Tsu}. A selfconsistent treatment of kaon-nucleon scattering
in the hadronic environment would be desirable in this context,
as it has been already
investigated for the antikaonic sector \cite{tolos}. Furthermore, a 
more realistic density and, in particular, energy dependence of the hadronic
relativistic mean field would be important \cite{NLD}.
However, in order to keep the discussions transparent, we do not expand on these
topics here, and use the same parametrization as in \cite{prassa1}.

\begin{figure}[t]
\unitlength1cm
\begin{picture}(8.,7.0)
\put(2.0,0.0){\makebox{\epsfig{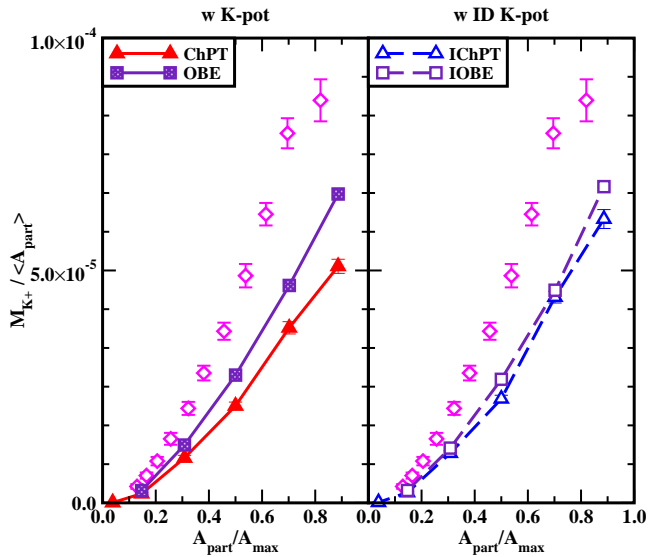}}}
\end{picture}
\caption{$K^{+}$ centrality dependence in Au+Au reactions at
$1$ AGeV incident energy.
Our theoretical calculations (as indicated) are compared with
KaoS data from \protect\cite{expa} (open diamonds).
}
\label{kpotcentr}
\end{figure}

In Fig.~\ref{kpotcentr}  the centrality dependence of the $K^{+}$ yields
for $Au+Au$-collisions is shown. First of all, all sets of data understimate
the data due to the strong density dependence of the inelastic cross section 
for $\Delta$-production, see also Ref. \cite{prassa1} and the discussion above. 
The in-medium dependence
of the kaon potentials is now more pronounced for the most central
$Au+Au$-collisions, see left panel in Fig.~\ref{kpotcentr}. Isospin effects
of both kaon potentials appear due to the asymmetry of the colliding
$Au$-systems, see right panel of the same figure. The effect of the inclusion of the
isovector potential turns out to be stronger in the $IChPT$ case
compared to the $IOBE$ one, since the in-medium energy for $K^{+}$ is
stronger affected in the $ChPT$ model, see again Fig.~\ref{kmass}. Thus, the
kaon yield is enhanced more strongly in the $IChPT$ transport calculations, and
comes closer to the $IOBE$ calculations, which are less sensitive to isospin
effects.

Collective kaon ($K^{+}$) flow (not shown here) has been also studied in the
framework of the different kaon potentials. All the calculations reproduce
the experimental flow \cite{FOPI_flows} reasonable well, however, the isospin
effects were only moderate. It would be more helpful to study {\it isospin}
strangeness flows, i.e., the difference of collective flows between $K^{0}$
and $K^{+}$, since the isovector part of the kaon potential affects $K^{+}$ and
$K^{0}$ differently. In conclusion, the evaluation scheme of the kaon-nucleon
potential within the effective $OBE$ approach seems to lead to a better
description of the experimental data. However, a definitive conclusion on
the question, which potential would be more realistic, cannot be drawn at this level
of study. A more systematic study on in-medium effects also on the level of
$NN$-scattering is in this context neccesary.

Thus at this level of the comparison between the theoretical
calculations and the data the $OBE$ framework describes the experiment better
than the $ChPT$ model. However, for a transparent
discussion specifically of the isospin effects on strangeness ratios both models of the
kaon-nucleon interaction will be investigated. A more promising observable to
study isovector effects may be given by the ratio
of the two isospin kaon states, e.g., the strangeness ratio $K^{0}/K^{+}$, since
ratios are expected to be independent on possible in-medium modifications of the
isoscalar medium. E.g., the influence of in-medium inelastic $NN\rightarrow N\Delta$
cross sections has been studied in \cite{prassa1} with the conclusion of a very
moderate dependence of the strangeness ratio on the density dependence of the
inelastic cross section.

\subsection{Strangeness ratio}

A crucial question is whether particle yield {\it ratios} are
influenced by the in-medium modifications of kaon potentials.
This is important particularly for kaons,
since ratios of particles with strangeness have been widely
used to determine the nuclear EoS at supra-normal density \cite{fuchs}.
More recently, the $(\pi^{-}/\pi^{+})$- and $(K^{0}/K^{+})$-ratios have
been proposed as a good probe to determine the high density behaviour of
 the isovector part of the nuclear mean field
\cite{ferini,ferini2,stoecker,bao}. We thus extend our previous studies
\cite{ferini2} by 
explicitly including both schemes of the kaon potential in the transport
calculations ($ChPT$ and $OBE$ models). In particular, we distinguish
between the different scenarios for the symmetry energy in the baryon sector, 
i.e., by neglecting the isovector potential ($NL$ case), by including only
the $\rho$-meson ($NL\rho$ case), and both, the $\rho$ and $\delta$ mesons
($NL\rho\delta$) in the isovector baryon and kaon potentials ($IChPT$ and $IOBE$
cases).

\begin{figure}[t]
\unitlength1cm
\begin{picture}(8.,7.0)
\put(2.0,0.0){\makebox{\epsfig{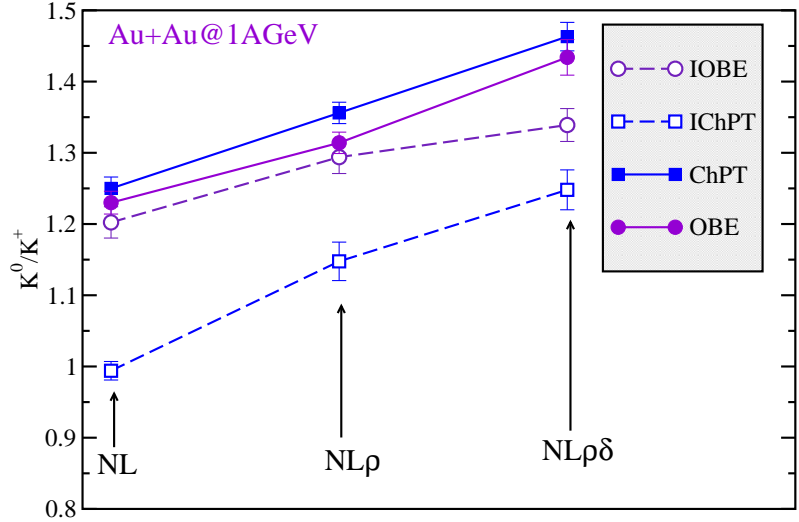}}}
\end{picture}
\caption{$K^{0}/K^{+}$ ratio for different parametrisations of the isovector
$EoS$ ($ NL,~NL\rho,~NL\rho\delta$) and for the kaon potential
($OBE$, $ChPT$, $IOBE$, $IChPT$). The considered reaction is a
central ($b=0$ fm) $Au+Au$ heavy-ion collision at $1~AGeV$ incident energy.
}
\label{ratios}
\end{figure}
Fig.~\ref{ratios} shows the strangeness ratio for the different
parametrisations of the isovector part of the nuclear $EoS$.
Calculations are also performed for the two approaches of the kaon
potential ($IChPT$ and $IOBE$ model).

First of all, an overall linear increase
of the ratio with respect to the stiffness of the symmetry energy is observed.
This feature is not trivial and reflects the interplay of isospin effects between
mean-field and collision dynamics at the threshold. As discussed in detail
in Ref. \cite{ferini2}, isovector mean-field and threshold effects act differently on 
the isospin kaon states $K^{+}$ and $K^{0}$, where the later feature governs the
strangeness dynamics at low intermediate energies near the kaon production threshold.

The sensitivity of the strangeness ratio on the different evaluations of the isospin
{\it independent} kaon potential ($ChPT$ and $OBE$ cases) is very moderate, as expected.
If the isovector part is neglected, the density behavior of the kaon potential is
the same for both isospin $K^{0}$ and $K^{+}$ states. Therefore the ratio should
be rather independent on the evaluation scheme of the kaon potential, as it is the case in
the transport calculations of Fig.~\ref{ratios}.

Taking explicitly account of the isospin effects in the kaon potentials ($IChPT$ and
$IOBE$ cases) the dynamical situation changes. The isospin effect in the $IChPT$ model
strongly decreases the strangeness ratio, however, the relative difference between
the various choices of the symmetry energy ($NL$, $NL\rho$ and $NL\rho\delta$) remains
stable. This is due to the fact that the isovector sector of the $IChPT$ kaon
self energies contain only the isospin Lorentz-scalar and Lorentz-vector densities
without additional parameters, e.g., meson-nucleon couplings \cite{prassa1}. This
feature is also shown in Fig.~\ref{kmass}, in which the isospin splitting is independent
on the stiffness choice of the symmetry energy. The decrease of the ratio in the $IChPT$
calculations is related to the fact that the in-medium energy becomes more stiffer
(softer) for $K^{0}$ ($K^{+}$). The situation in the $IOBE$ calculations is similar,
but the isospin effects are less pronounced. The suppression of the kaon ratio in the
transport calculations due to the isospin effects is now weak, due to the softer
character of the $(I)OBE$ kaon potential, compared to the $(I)ChPT$ model. However,
the relative difference between the different models of the symmetry energy is
slightly modified. When isospin effects are accounted for, the $IOBE$ self energies
explicitly contain the meson-nucleon couplings, which are different for each model of the
symmetry energy, see also again Fig.~\ref{kmass}. This leads to the observed different
slope of the kaon ratio in the $IOBE$ transport calculations, however, the effect is rather
moderate. It is interesting to note that the situation in a chemical and thermal
equilibrated (closed) system, e.g., in studying strangeness production in a box with
periodic boundary conditions, is rather different. The isospin effects on the strangeness
ratio are very pronounced in box calculations (see Fig.~$8$ in Ref. \cite{ferini}).
However, the isospin effect is considerably reduced in an open system in heavy-ion
collisions due to, e.g., isospin transparency which decreases the local asymmetry
of the kaon production source.

In summary, the consideration of the isospin effects on the level of the kaon-nucleon
interaction leads to modifications of the strangeness ratio on the absolute value,
but not to significant changes with respect to the stiffness of the symmetry energy.
In the $IChPT$ and $IOBE$ approaches a sensitivity to the symmetry energy on the
level of $\sim 25\%$ and $\sim 15\%$ survives. We thus conclude that the strangeness
ratio is a very promising observable to determine the isospin dependence of the
underlying kaon-nucleon interaction and particularly the isospin part of the
nuclear EoS.

\section{Conclusions}
We have investigated the role of different kaon potentials and, in
particular, their explicit isospin dependence in asymmetric nuclear matter.
The kaon potentials have been evaluated in two schemes, one following a Chiral
Perturbation approach and a second one within the framework of an effective
One-Boson-Exchange picture.

We have studied in detail the effects of various isospin dependent kaon potentials
on kaon dynamics in heavy-ion collisions within a covariant transport equation.
The isospin effects turns out to be less sensitive to absolute kaon yields, but
more pronounced to the strangeness $K^{0}/K^{+}$ ratio. We thus focused our attention
to the $K^{0}/K^{+}$ ratio in collisions between heavy and
isospin-asymmetric systems ($Au+Au$) at incident energies below the kaon production
threshold. Different evaluation schemes of the kaon potential and its isovector
part lead to different results on the strangeness ratio. The isospin dependence
changes this ratio on an absolute level. As an interesting result, however, the
relative dependence of the strangeness ratio to the stiffness of the isovector EoS
remains stable.

At this level of investigation we conclude that the strangeness ratio is an
important observable to determine the high density behavior of the symmetry
energy. Future experiments with exotic radioactive beams, as they are planned
in new experimental facilities at MSU and at GSI (FAIR project), will be very
helpful to resolve the theoretical uncertainties of the symmetry energy at high
baryon densities.

{\bf Acknowledgements}\\
We acknowledge very fruitful discussions with A.B. Larionov, M. Kaskulov,
H. Lenske. We would like to particularly thank M. Kaskulov and
H. Lenske for a carefully reading of the manuscript and further
suggestions. This work is supported by BMBF 
and by the DFG Cluster of Excellence {\it Origin and structure of the Universe}.
One of the authors (V.P.)
acknowledges support by the Hellenic State Scholarships Foundation (I.K.Y.).


\end{document}